\documentclass[journal]{IEEEtran}

\usepackage{algorithmic}
\usepackage{textcomp}
\def\BibTeX{{\rm B\kern-.05em{\sc i\kern-.025em b}\kern-.08em
    T\kern-.1667em\lower.7ex\hbox{E}\kern-.125emX}}

\usepackage{graphicx}
\usepackage[font=footnotesize]{subfig}
\usepackage{float}
\usepackage{color}
\usepackage{bbm}
\usepackage{dsfont}
\usepackage{cite}
\usepackage{balance}
\usepackage{url}
\usepackage{times}
\usepackage{amsbsy}
\usepackage[T1]{fontenc}
\usepackage[latin9]{inputenc}
\usepackage{tablefootnote}
\usepackage{threeparttable}
\usepackage{amsmath,amssymb,amsfonts}
\usepackage{subfig}
\usepackage[font=small]{caption}
\usepackage{multirow}
\renewcommand\IEEEkeywordsname{Index Terms}

\usepackage{resizegather}

\usepackage{suffix}
\usepackage{mathtools}

\DeclarePairedDelimiterX\MeijerM[3]{\lparen}{\rparen}%
{\,#3\delimsize\vert\begin{smallmatrix}#1 \\ #2\end{smallmatrix}}

\newcommand\MeijerG[8][]{%
  G^{\,#2,#3}_{#4,#5}\MeijerM[#1]{#6}{#7}{#8}}

\WithSuffix\newcommand\MeijerG*[7]{%
  G^{\,#1,#2}_{#3,#4}\MeijerM*{#5}{#6}{#7}}

\newcommand{\subparagraph}{}
\usepackage{titlesec}

\titlespacing*{\subsubsection}
{0pt}{.086cm}{.1cm}

\usepackage{romannum}

\begin{document}

\bstctlcite{IEEEexample:BSTcontrol}

\title{Secure Outage Analysis of FSO Communications Over Arbitrarily Correlated M{\'a}laga Turbulence Channels}

\author{
Yun Ai, \IEEEmembership{Member, IEEE}, Aashish Mathur, \IEEEmembership{Member, IEEE}, Long Kong, and  Michael Cheffena

\thanks{Y. Ai and M. Cheffena are with the Norwegian University of Science and Technology (NTNU), A. Mathur is with the Indian Institute of Technology Jodupur, L. Kong is with the University of Luxembourg.}
}

{}


\maketitle

\begin{abstract}
In this paper, we analyze the secrecy outage performance for more realistic eavesdropping scenario of free-space optical (FSO) communications, where the main and wiretap links are correlated. The FSO fading channels are modeled by the well-known M{\'a}laga distribution. Exact expressions for the secrecy performance metrics such as secrecy outage probability (SOP) and probability of the non zero secrecy capacity (PNZSC) are derived, and asymptotic analysis on the SOP is also conducted. The obtained results reveal useful insights on the effect of channel correlation on FSO communications. Counterintuitively, it is found that the secrecy outage performance demonstrates a non-monotonic behavior with the increase of correlation. More specifically, there is an SNR penalty for achieving a target SOP as the correlation increases within some range. However, when the correlation is further increased beyond some threshold, the SOP performance improves significantly.
\end{abstract}


\IEEEpeerreviewmaketitle

\section{Introduction \label{sec:introduction}}

\IEEEPARstart{P}{hysical} layer security (PLS) has recently appeared as a complement to the conventional encryption techniques to enhance the communication secrecy \cite{barros2006secrecy}. It is shown that fading, which is usually considered as a negative factor in terms of reliability, can be used to increase the communication security  against eavesdropping \cite{barros2006secrecy, ai2019secrecyisl, pan2017secure}. In practice, a number of scenarios (e.g., the proximity of the legitimate receiver and eavesdropper, antenna deployments, and the scattering conditions, etc.) may lead to the spatial fading correlation between the fading links in the PLS setup \cite{alexandropoulos2018secrecy}. This motivates the research on the impact of the fading channel correlation on the secrecy performance \cite{alexandropoulos2018secrecy, jeon2011bounds, ferdinand2013physical, liu2013outage, pan2016physical, mathur2019secrecycorrelated}.




The secrecy outage probability (SOP) performance was studied for correlated composite Nakagami-$m$/Gamma fading channels in \cite{alexandropoulos2018secrecy}. The impact of spatial correlation on the secrecy capacity was quantified in \cite{jeon2011bounds} assuming channel state information (CSI) of both the legitimate receiver and eavesdropper available at the transmitter. In \cite{ferdinand2013physical}, the secrecy performance of Multiple-Input Multiple-Output (MIMO) wiretap channels with orthogonal space-time block codes was studied by considering the signal correlation due to antenna elements proximity. By assuming that both links of the Wyner's wiretap model are arbitrarily log-normally distributed, the asymptotic SOP performance was studied in \cite{liu2013outage}. The secrecy performance over correlated log-normal and $\alpha$-$\mu$ fading channels were investigated in \cite{pan2016physical} and \cite{mathur2019secrecycorrelated}, respectively. By considering significant beamforming and thus ignoring the effect of small-scale fading, the secrecy performance for an mmWave ad-hoc network was studied in \cite{zhu2017secure}. The PLS performance of an intelligent reflecting surfaces (RIS) aided terahertz (THz) communication system is investigated in \cite{qiao2020secure}.


Recently, secrecy analysis for optical communications has been a research topic for both visible light \cite{pan2017securecl} and free-space optical (FSO) communications \cite{lopez2015physical, sun2016physical, saber2017secure, monteiro2018maximum}. In \cite{lopez2015physical}, the expression for the Probability of the Non-zero Secrecy Capacity (PNZSC) was derived for the FSO communications by ignoring the FSO link fading and only considering the atmospheric turbulence. The PLS of a line-of-sight (LOS) FSO link using orbital angular momentum (OAM) multiplexing was studied in \cite{sun2016physical}. The secrecy performance of FSO communication was studied in \cite{saber2017secure} by assuming both the main and wiretap links following independent M{\'a}laga distributions. The secrecy throughput of the coherent FSO communication in the presence of a MIMO multi-apertures eavesdropper was studied for the Gamma-Gamma fading channel conditions in \cite{monteiro2018maximum}.

It is clear from above works on the secrecy performance of FSO communications and the references therein that, the fading correlation between the main and wiretap channels have not been considered. The relatively good directionality of the wireless optical transmission makes it inherently more secure than the radio-frequency (RF) transmission \cite{lopez2015physical, sun2016physical, saber2017secure, monteiro2018maximum, lei2018secrecytcomm}. However, this good directionality of wireless optical signals also means that the eavesdropper needs to be placed close to the legitimate receiver for the purpose of eavesdropping, which implies that the main and wiretap links are more likely to experience correlated fading in FSO communications.

Motivated by the latest advances in the PLS analysis on FSO communication and aiming at investigating the PLS performance of FSO communications under more realistic conditions, we study in this paper the secrecy performance of FSO communications  over arbitrarily correlated M{\'a}laga fading channels. The choice of M{\'a}laga distribution as the investigated statistical model is justified by its applicability to all atmospheric turbulence regimes as well as its generality. For instance, the M{\'a}laga model encompasses some of the most widely used distributions such as log-normal, exponential, and Gamma-Gamma, etc. \cite{jurado2012further}. The main contributions of this paper are: (\romannum{1}) We analyze the secrecy outage performance of a more realistic eavesdropping scenario for FSO communications, where the main and wiretap links are correlated; (\romannum{2}) The effects of correlation on SOP and PNZSC are evaluated, which demonstrates that correlation can potentially be utilized for SOP performance enhancement; (\romannum{3}) Different from previous works on secrecy analysis of FSO communications assuming independent fading conditions \cite{lopez2015physical, sun2016physical, saber2017secure}, we derive exact expression for SOP rather than lower bound of SOP. The lower bounds of SOP are tight only in some cases \cite{lei2018secrecytcomm}; and (\romannum{4}) The diversity analysis is conducted  and useful insights on asymptotic slope of the SOP curve is obtained.




\emph{Notations}: $[x]^{+} = \max(x, 0)$, $\Gamma(\cdot)$ is the Gamma function \cite[p.~759]{prudnikov1986integrals}. $G^{m,n}_{p,q}\!(\cdot)$ is the Meijer G-function \cite[Eq.~(8.2)]{prudnikov1986integrals}, $U(\cdot, \cdot, \cdot)$ is the Kummer hypergeometric function \cite[p.~793]{prudnikov1986integrals}, and ${}_{p}F_{q}{(\cdot; \cdot; \cdot)}$ is the hypergeometric function \cite[p.~333]{prudnikov1986integrals}.



\section{Channel and System Models \label{sec:channel_system_model}}

\begin{figure}[t!]
\centering
  \includegraphics[width=0.96\linewidth,keepaspectratio,angle=0]{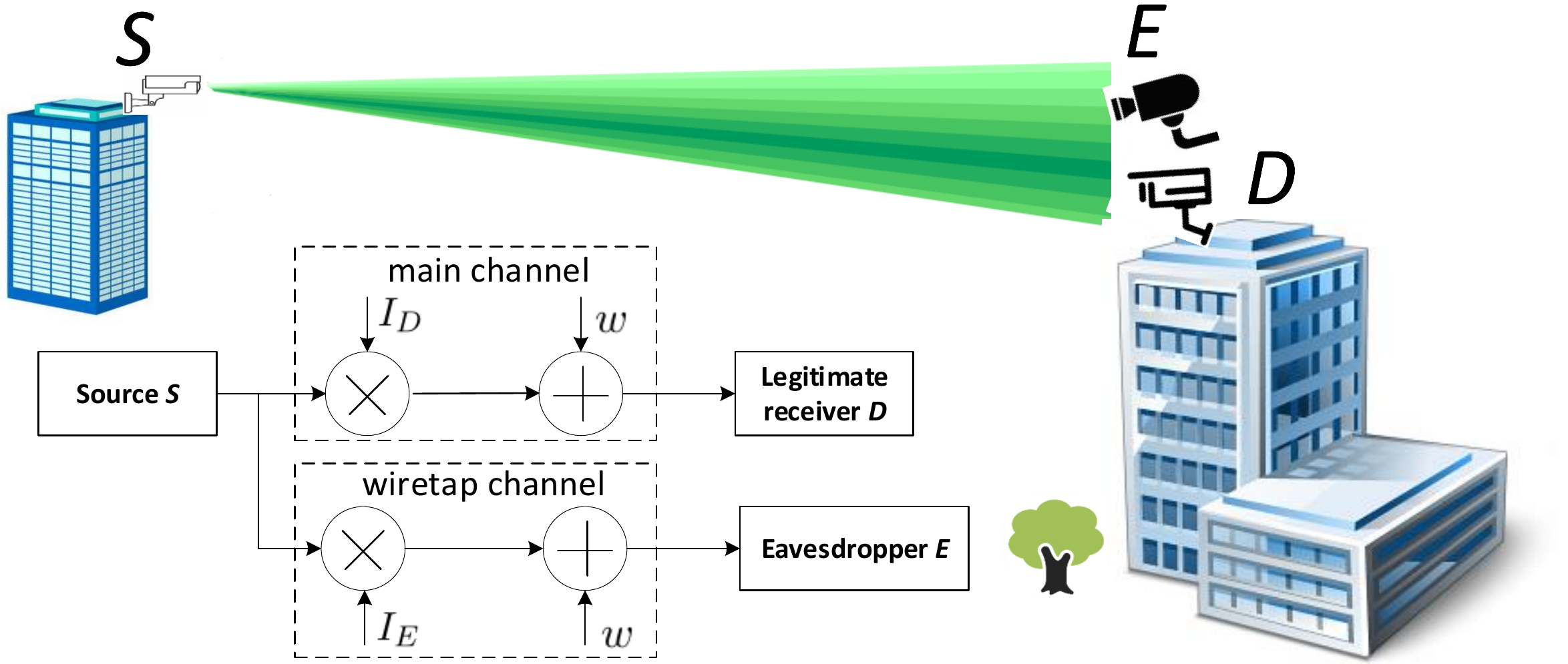}
  \caption{Investigated PLS scenario for FSO communications.}
  \label{fig:system_model}
\end{figure}

In this paper, we consider an intensity modulation/direct detection (IM/DD) based FSO system. The fading behaviors of the FSO links are characterized by the generalized M{\'a}laga model, which takes into account three components: the LOS component $L_{S}$, the component $L_{S}^{C}$ that is coupled to $L_{S}$ and quasi-forward scattered by the eddies on the propagation axis, and the component $L_{S}^{G}$ resulting from the energy that is scattered by off-axis eddies \cite{priyadarshani2018outage}.


The studied classic Wyner's wiretap model under passive eavesdropping is illustrated in Fig. \ref{fig:system_model}. The legitimate source \emph{S} transmits signals to the legitimate destination node \emph{D} over the main channel. The eavesdropper \emph{E} attempts to intercept the information by decoding its received signal from the wiretap channel. The main and wiretap channels follow M{\'a}laga fading and are assumed to be arbitrarily correlated due to close proximity of the nodes \emph{D} and \emph{E} or similarity of the scatterers around them. Moreover, we assume that both channels undergo ergodic block fading, i.e., the channel coefficients remain constant during the transmission of one block of codewords and vary independently from one block period to the next one.

The regenerated electrical signal at the FSO receiver nodes \emph{D} and \emph{E}, respectively, can be expressed as \cite{jurado2012further, alexandropoulos2018secrecy}
\begin{align}
y_{x} = & \eta I_{\!x} s + w = \eta Y_{\!x} X_{\!x} s + w,  \label{eq:recd_signal}
\end{align}
where $x=\{D, E\}$, $s$ is the transmitted symbol with unit energy, the random variables (RVs) $I_{\!D}$ and $I_{\!E}$ represent the received signal irradiance affected by atmospheric turbulence at the corresponding receiver aperture, $\eta$ is the optical-to-electrical conversion coefficient, $w$ is the additive white Gaussian noise (AWGN) with power spectral density $\frac{N_{0}}{2}$, which, without loss of generality, is assumed to be the same for both channel links. In (\ref{eq:recd_signal}), $Y_{x}$ and $X_{x}$, $x \in \{D, E\}$, represent the small-scale and large-scale fluctuations, respectively  \cite{jurado2012further}. As in \cite{alexandropoulos2018secrecy}, we consider the realistic scenario that the turbulent flow of the large-scale eddies induces the correlation while the small-scale fluctuation is assumed to be independent between \emph{D} and \emph{E}. The instantaneous signal-to-noise ratio (SNR) between \emph{S} and $x$, $x \in \{D, E\}$, can be written as $\gamma_{\scriptscriptstyle{S} \! x} = \frac{ \eta^{2} Y_{\!x}^{2} X_{\!x}^{2} }{N_{0}}$.

Let us denote the instantaneous SNRs for the \emph{S}-\emph{D} and \emph{S}-\emph{E} links $\gamma_{1}$ and $\gamma_{2}$, respectively, for simplicity. The joint probability density function (PDF) of the $\gamma_{1}$ and $\gamma_{2}$ over the considered arbitrarily correlated M{\'a}laga fading channel can be obtained from \cite[Eq.~(7)]{priyadarshani2018outage}, after some algebra, as follows:
\begin{align}
& f_{ \gamma_{1}, \gamma_{2} }( \gamma_{1}, \gamma_{2} ) =  \sum_{t=0}^{\infty} \mathcal{F}_{t}    \cdot    \prod_{p=1}^{2} \biggl[ \frac{ 1 }{ 2 \sqrt{ \gamma_{p} \mu_{p} } }  \sum_{k=1}^{\beta} (-1)^{k-1}   \cdot \!  {  \beta - 1  \choose  k-1   }   \nonumber \\ &  \cdot \frac{ 1 }{ (k - 1)! } \!  \left( \frac{ \beta \Omega ( 1 - \rho^{2} ) \sqrt{ \gamma_{p} } }{ ( \xi \beta + \Omega_{1} )  \sqrt{ \mu_{p} } }    \right)^{ \! \frac{ \alpha + t - k }{ 2 } }  \!  \left(   - \frac{ \Omega_{1} \sqrt{ \gamma_{p} } }{ ( \xi \beta + \Omega_{1} ) \xi \sqrt{ \mu_{p} } }    \right)^{ \!\! k - 1 }   \nonumber \\  &   \cdot   \MeijerG*{2}{0}{0}{2}{ - }{ \frac{ \alpha + t - k }{ 2 }, - \frac{ \alpha + t - k }{ 2 }  }{ \frac{ \beta }{ \Omega ( 1 - \rho^{2} ) ( \xi \beta + \Omega_{1} ) }  \sqrt{ \frac{ \gamma_{ p } }{ \mu_{p} } } }  \biggr],
\label{eq:joint_pdf}
\end{align}
where
\begin{align}
\mathcal{F}_{t} = \xi^{ 2(\beta - 1) } \! \cdot \!  \left( \frac{ \beta }{ \xi \beta + \Omega_{1} } \right)^{ 2 \beta } \!\! \cdot \!  \frac{ ( 1 - \rho^{2} )^{ - \alpha - 2 t }  \cdot \rho^{2t} }{ \Gamma( \alpha )   \Gamma(t)  \Gamma( \alpha \! + \! t ) \cdot \Omega^{ 2 \alpha + 2 } },
\label{eq:joint_pdf_app}
\end{align}
and $\rho \in [0, 1)$ denotes the correlation factor between the M{\'a}laga fading channels, $\mu_{p} = \mathrm{E}\{ \gamma_{p} \}, p\in\{1, 2\}$, represents the average SNR of the corresponding link, the parameter $\alpha$ describes the fading severity due to atmospheric turbulence, $\beta$ is a natural number related to the effective number of small-scale cells, $\xi = 2 b_{0} ( 1 - \delta )$ is the average power of the scattering component with $2 b_{0}$ being the average power of the total scatter components and $\delta$ ($0 < \delta < 1$) being the amount of scattering power coupled to LOS component,  $\Omega$ denotes the average power of the large-scale fluctuation, $\Omega_{1} = \Omega' + 2 b_{0} \delta + 2 \sqrt{ 2 b_{0} \delta \Omega' }\cos( \phi_{A} - \phi_{B}  ) $, where $\Omega'$ is the average power of the LOS component, $\phi_{A}$ and $\phi_{B}$ denote the deterministic phases of the LOS component and the scatterers coupled to the LOS component, respectively \cite{priyadarshani2018outage, jurado2011unifying}.


\section{Secrecy Performance Analysis \label{sec:secrecy_analysis}}

\subsection{Secrecy Outage Probability (SOP) \label{subsec:sop_analysis}}

The secrecy rate refers to the maximum achievable rate the legitimate main channel can achieve in secrecy under eavesdropping. The instantaneous secrecy rate $C_{s}$ of the considered wiretap model is defined as \cite{ai2018physical}
\begin{align}
C_{s}(\gamma_{1}, \gamma_{2}) = \left[\ln(1+\gamma_{1}) - \ln(1+\gamma_{2}), 0 \right]^{+},
\label{eq:inst_asc_def}
\end{align}
where $\ln(1+\gamma_{1})$ and $\ln(1+\gamma_{2})$ are the instantaneous capacities of the main link $S$-$D$ and wiretap link $S$-$E$ as illustrated in Fig. \ref{fig:system_model}, respectively.

Under passive eavesdropping attack, the legitimate source node $S$ and receiver $D$ have no CSI on eavesdropper $E$'s channel. In this case, the source node $S$ cannot adapt the coding scheme to $E$'s channel condition, but resorts to set the secrecy rate to a constant target rate $R_{s}$. When $C_{s} > R_{s}$, perfect secrecy can be achieved. Otherwise, when $C_{s} \leq R_{s}$, secrecy will be compromised and secrecy outage occurs, the probability of which can be evaluated by the security metric SOP \cite{barros2006secrecy}. The SOP is mathematically expressed as \cite{ai2019secrecyel}
\begin{align}
P_{o}  =  &  \mathrm{Pr}\left[ C_{s}(  \gamma_{1}, \gamma_{2}  ) \leq  R_{s} \right]  =  \mathrm{Pr}\left[  \gamma_{1}  \leq \Theta \gamma_{2} + \Theta - 1  \right]   \nonumber  \\
       =  &  \int_{0}^{ \infty }  \int_{ 0 }^{ ( 1 + \gamma_{2} ) \Theta - 1 }  f_{ \gamma_{1}, \gamma_{2} }( \gamma_{1}, \gamma_{2} )  \,  d \gamma_{1}   d\gamma_{2},
\label{eq:sop_def}
\end{align}
where $\Theta=\exp(R_s) \geq 1$.


Substituting (\ref{eq:joint_pdf}) into (\ref{eq:sop_def}), the SOP can be expressed as
\begin{align}
P_{o}  = & \sum_{t=0}^{\infty} \mathcal{F}_{t}    \cdot    \prod_{p=1}^{2} \biggl[ \frac{ 1 }{ 2 \sqrt{ \mu_{p} } }  \sum_{k=1}^{\beta} (-1)^{k-1}   \cdot   {  \beta - 1  \choose  k-1   }    \cdot \frac{ 1 }{ (k - 1)! }   \nonumber \\ &   \cdot  \left( \frac{ \beta \Omega ( 1 - \rho^{2} )  }{ ( \xi \beta \! + \! \Omega_{1} )  \sqrt{ \mu_{p} } }    \right)^{ \! \frac{ \alpha + t - k }{ 2 } } \!\!\! \cdot \! \left(   - \frac{ \Omega_{1}  }{ ( \xi \beta \! + \! \Omega_{1} ) \xi \sqrt{ \mu_{p} } }    \right)^{ \!\! k - 1 }  \cdot  \mathcal{I}_{1}  \biggr]  ,
\label{eq:sop_def1}
\end{align}
where $\mathcal{I}_{1}$ is the double integral expressed as follows:
\begin{align}
\mathcal{I}_{1}  =  \int_{0}^{\infty}   \gamma_{2}^{ \frac{ \alpha + t + k - 4 }{ 4 } }    \MeijerG*{2}{0 }{0}{2 }{  -   }{   \mathcal{K}_{1}  }{ \frac{ \beta ( 1  -  \rho^{2} )^{-1} \sqrt{\gamma_{ 2 }} }{ \Omega ( \xi \beta  +  \Omega_{1} ) \sqrt{\mu_{ 2 }} }  }  \cdot \mathcal{I}_{1a} \, d\gamma_{2}.
\label{eq:sop1}
\end{align}

In (\ref{eq:sop1}), $\mathcal{I}_{1a}$ is given as
\begin{align}
\mathcal{I}_{1a}  =  \int_{0}^{ ( 1 + \gamma_{2} ) \Theta - 1 }  \gamma_{1}^{ \frac{ \alpha + t + k - 4 }{ 4 } } \cdot \MeijerG*{2}{0 }{0}{2 }{ -   }{  \mathcal{K}_{1}  }{ \frac{ \mathcal{C} \sqrt{\gamma_{ 1 }} }{  \sqrt{\mu_{ 1 }} }  } \, d \gamma_{1} ,
\label{eq:sop1a}
\end{align}
where $\mathcal{C} = \frac{ \beta  }{ \Omega ( \xi\beta + \Omega_{1} ) ( 1 - \rho^{2} )  }$ and $\mathcal{K}_{1} = (\frac{ \alpha + t - k }{ 2 }, - \frac{ \alpha + t - k }{ 2 })$. With the assistance of property \cite[Eq.~(2.24.2.2)]{prudnikov1986integrals}, the integral $\mathcal{I}_{1a}$ can be solved as
\begin{align}
\mathcal{I}_{1a} \!  = \! \frac{ [ ( 1  +  \gamma_{2} ) \Theta \! - \! 1 ]^{  \frac{ \alpha  +  t + k }{4} } }{ 2 \pi   } \!  \cdot \! \MeijerG*{4}{1 \!\!}{1}{5 \!\!}{ \! 1  -  \frac{ \alpha + t + k }{ 4 }  }{\! \mathcal{K}_{2}   }{\!\! \frac{ \mathcal{C}^{2}   [ ( 1 \! + \! \gamma_{2} ) \Theta \! - \! 1 ]   }{ 16   \mu_{1}   }  } ,
\label{eq:sop2}
\end{align}
where $\mathcal{K}_{2} \!\! = \!\! ( \frac{ \alpha + t - k }{ 4 }, \frac{ \alpha + t - k + 2 }{ 4 }, - \frac{ \alpha + t - k }{ 4 }, - \frac{ \alpha + t - k - 2 }{ 4 } , - \frac{ \alpha + t + k }{ 4 } )$.



Substituting (\ref{eq:sop2}) into (\ref{eq:sop1}), the exact expression for SOP can be given as the following one-fold integral:
\begin{align}
\mathcal{I}_{1}  = &  \int_{0}^{\infty}    \frac{ \gamma_{2}^{ \frac{ \alpha + t + k - 4 }{ 4 } }   }{ 2 \pi [ ( 1  +  \gamma_{2} ) \Theta  -  1 ]^{ - \frac{ \alpha  +  t + k }{4} }  }   \cdot   \MeijerG*{2}{0 }{0}{2 }{  -   }{  \mathcal{K}_{1}  }{ \frac{ \mathcal{C} \sqrt{\gamma_{ 2 }} }{  \sqrt{\mu_{ 2 }} }  }  \nonumber \\ &  \cdot  \MeijerG*{4}{1 }{1}{5 }{  1  -  \frac{ \alpha + t + k }{ 4 }  }{ \mathcal{K}_{2}   }{ \frac{ \mathcal{C}^{2}   [ ( 1  +  \gamma_{2} ) \Theta  -  1 ]   }{ 16  \mu_{1}  }  } \, d\gamma_{2}.
\label{eq:sop3}
\end{align}

The resulting integral in (\ref{eq:sop3}) cannot in general be solved in closed form when the target secrecy rate $R_{s} > 0$. Therefore, we present an efficient and accurate numerical approach to evaluate the SOP. From the equalities \cite[Eq.~(14)]{adamchik1990algorithm} and \cite[Eq.~(9.238.3)]{jeffrey2007table}, the following equality holds:
\begin{align}
\MeijerG*{2}{0\! }{0}{2 \!}{ -  \!\! }{ \!\! \frac{v}{2}, -\frac{v}{2}  \!\! }{\!\!  \frac{ \mathcal{C} \sqrt{\gamma}_{2}  }{  \sqrt{ \mu_{2} } }  } = \frac{ \mathcal{A}_{1} }{ \mathcal{A}_{2} } \cdot U\!\!\left( \frac{1}{2} + v, 1 + 2v; 4\left(  \frac{ \mathcal{C}^{2} \gamma_{2} }{ \mu_{2} } \right)^{ \frac{ 1 }{ 4 } } \right),
\label{eq:sop3a}
\end{align}
where $v \! = \!  \alpha  +  t  -  k$, $\mathcal{A}_{1}\!\! = \!\! 2^{2v+1} \!  \sqrt{\pi}  \bigl(\! \frac{ \mathcal{C}^{2} \gamma_{2} }{ \mu_{2} } \!\bigr)^{ \frac{v}{4} }$, and $\mathcal{A}_{2}\! =\! e^{ 2 \bigl(\! \frac{ \mathcal{C}^{2} \gamma_{2} }{ \mu_{2} }  \!\bigr)^{ \frac{1}{4} } }$.

Utilizing the above relation (\ref{eq:sop3a}) in (\ref{eq:sop3}), and then making the following change of RVs: $\frac{ \mathcal{C} \sqrt{\gamma}_{2}  }{  \sqrt{ \mu_{2} } } = \frac{ z^{4} }{ 4 }$, leads to
\begin{align}
\mathcal{I}_{1} \! = \! &   \int_{0}^{\infty}  \!\!  \frac{ e^{-z^{2}} 2^{v+3} z^{2v-1}  ( \frac{ \mu_{2} z^{8} }{ 16 \mathcal{C}^{2} } )^{ \frac{ \alpha + t + k }{ 4 } }  }{ \sqrt{\pi}  [ ( 1 + \frac{ \mu_{2} z^{8} }{ 16 \mathcal{C}^{2} } )\Theta - 1  ]^{ - \frac{ \alpha + t + k }{ 4 }} }    \cdot   U\!\!\left( \frac{1}{2} + v, 1 + 2v; 2 z^{2} \right)  \nonumber \\ &  \cdot  \MeijerG*{4}{1 }{1}{5 }{  1  -  \frac{ \alpha + t + k }{ 4 }  }{ \mathcal{K}_{2}   }{ \frac{ \mathcal{C}^{2}   [ ( 1  +  \frac{ \mu_{2} z^{8} }{ 16 \mathcal{C}^{2} } ) \Theta  -  1 ]   }{ 16  \mu_{1}  }  } \, d z .
\label{eq:sop3b}
\end{align}

Then, the integral $\mathcal{I}_{1}$ can be efficiently and accurately evaluated with the modified Gauss-Chebyshev quadrature technique \cite{steen1969gaussian} as in (\ref{eq:sop4}) at the top of next page. In (\ref{eq:sop4}), $w_{\iota}$ and $s_{\iota}$, ($\iota = 1,\dots, L$), are respectively the weights and abscissas of the $L$-order polynomial, as detailed in \cite{steen1969gaussian}. Substituting (\ref{eq:sop4}) into (\ref{eq:sop_def1}), the expression for the SOP is obtained.

\begin{figure*}[t!]
\begin{gather}
\mathcal{I}_{1} \! = \! \sum_{\iota = 1}^{L} w_ {\iota} \cdot \frac{   2^{v+3} s_{\iota}^{2v-1}  ( \frac{ \mu_{2} s_{\iota}^{8} }{ 16 \mathcal{C}^{2} } )^{ \frac{ \alpha + t + k }{ 4 } }  }{ \sqrt{\pi}  [ ( 1 + \frac{ \mu_{2} s_{\iota}^{8} }{ 16 \mathcal{C}^{2} } )\Theta - 1  ]^{ - \frac{ \alpha + t + k }{ 4 }} }    \cdot   U\!\!\left( \frac{ \alpha + t - k + 1 }{ 2 }, \alpha + t - k + 1; 2 s_{\iota}^{2} \right)  \cdot  \MeijerG*{4}{1 }{1}{5 }{  1  -  \frac{ \alpha + t + k }{ 4 }  }{ \mathcal{K}_{2}    }{ \frac{ \mathcal{C}^{2}   [ ( 1  +  \frac{ \mu_{2} s_{\iota}^{8} }{ 16 \mathcal{C}^{2} } ) \Theta  -  1 ]   }{ 16  \mu_{1}  }  }.
\label{eq:sop4}
\end{gather}
 \vspace*{-22.6pt}
\end{figure*}

\subsection{Probability of the Non-zero Secrecy Capacity (PNZSC) \label{subsec:pnzsc_analysis}}

The PNZSC is a fundamental security benchmark to characterize the existence of secrecy capacity. By definition, the PNZSC is written as the probability that the instantaneous secrecy rate is greater than zero, i.e., \cite{kong2019intercept}
\begin{align}
P_{1}  & =  \mathrm{Pr}\left[ C_{s}(  \gamma_{1}, \gamma_{2}  ) \geq  0 \right]  =  \mathrm{Pr}\left[  \gamma_{1}  \geq  \gamma_{2}  \right]   \nonumber  \\
       & =  1 - \int_{0}^{ \infty }  \int_{ 0 }^{ \gamma_{2} }  f_{ \gamma_{1}, \gamma_{2} }( \gamma_{1}, \gamma_{2} )  \,  d \gamma_{1}   d\gamma_{2} .
\label{eq:pnzsc_def}
\end{align}


Next, we derive a novel and exact analytical expression for the PNZSC over arbitrarily correlated M{\'a}laga fading links. By setting $\Theta = 1$ (i.e., $R_{s} = 0$) in (\ref{eq:sop1}), we have the following expression of PNZSC:
\begin{align}
P_{1}  = & 1 - \sum_{t=0}^{\infty} \mathcal{F}_{t}    \cdot    \prod_{p=1}^{2} \biggl[ \frac{ 1 }{ 2 \sqrt{ \mu_{p} } }  \sum_{k=1}^{\beta} (-1)^{k-1}  \! \cdot  \!  {  \beta - 1  \choose  k-1   }  \!  \cdot \! \frac{ 1 }{ (k - 1)! }   \nonumber \\ &   \cdot  \left( \frac{ \beta \Omega ( 1 - \rho^{2} )  }{ ( \xi \beta \! + \! \Omega_{1} )  \sqrt{ \mu_{p} } }    \right)^{ \! \frac{ \alpha + t - k }{ 2 } } \!\!\! \cdot \! \left(   - \frac{ \Omega_{1}  }{ ( \xi \beta \! + \! \Omega_{1} ) \xi \sqrt{ \mu_{p} } }    \right)^{ \!\! k - 1 }   \cdot  \mathcal{I}_{2}  \biggr]  ,
\label{eq:pnzsc_def1}
\end{align}
where
\begin{align}
\mathcal{I}_{2} \!  = \!\!\!  \int_{0}^{\infty}  \!\!  \frac{ \gamma_{2}^{ \frac{ \alpha + t + k  }{ 2 } - 1 }  }{ 2 \pi }  \!  \cdot \!   \MeijerG*{2}{0 }{0}{2 }{  -  \!\! }{  \mathcal{K}_{1}  \!\!  }{ \!\! \frac{ \mathcal{C} \sqrt{\gamma_{ 2 }} }{  \sqrt{\mu_{ 2 }} }  }  \!\! \cdot \!  \MeijerG*{4}{1 }{1}{5 }{  1  -  \frac{ \alpha + t + k }{ 4 }  \!\! }{ \mathcal{K}_{2}  \!\! }{ \!\! \frac{ \mathcal{C}^{2}   \gamma_{2}    }{ 16  \mu_{1}  }  }  d\gamma_{2}  .
\label{eq:pnzsc1}
\end{align}


Utilizing the equality \cite[Eq.~(2.24.1.1)]{prudnikov1986integrals} in (\ref{eq:pnzsc1}) leads to
\begin{align}
\mathcal{I}_{2}  = &  \frac{ 2^{ 2( \alpha + t + k - 1 ) }  \cdot  \mu_{2}^{ \frac{ \alpha + t + k }{2}   }}{  \pi^{2}  \cdot  \mathcal{C}^{ \alpha + t + k } }   \cdot   \MeijerG*{4}{5  }{ 5 }{ 5  }{ \mathcal{K}_{3}   }{ \mathcal{K}_{2}  }{    \frac{ \mu_{2} }{ \mu_{1} }  }  ,
\label{eq:pnzsc2}
\end{align}
where $\mathcal{K}_{3} = ( \frac{ 4 - \alpha - t - k }{ 4 }, \frac{ 2 - \alpha - t + k }{ 4 } , \frac{ 4 - \alpha - t + k }{ 4 } , \frac{ 2 +\alpha + t - k }{ 4 } , \frac{ 4 +\alpha + t - k }{ 4 } )$.


Finally, substituting (\ref{eq:pnzsc2}) into (\ref{eq:sop_def1}), we obtain the exact expression for the PNZSC in (\ref{eq:pnzsc3}) at the top of next page.


\begin{figure*}[t!]
\begin{gather}
P_{1}  =  1 - \sum_{t=0}^{\infty} \frac{ \mathcal{F}_{t} }{ 4 \pi^{2} }    \cdot    \prod_{p=1}^{2} \biggl[ \frac{ 1 }{ 2 \sqrt{ \mu_{p} } }  \sum_{k=1}^{\beta} (-1)^{k-1}   \cdot   {  \beta - 1  \choose  k-1   }    \cdot \frac{ 1 }{ (k - 1)! }     \cdot  \left( \frac{ \beta \Omega ( 1 - \rho^{2} )  }{ ( \xi \beta  +  \Omega_{1} )  \sqrt{ \mu_{p} } }    \right)^{  \frac{ \alpha + t - k }{ 2 } }  \cdot  \left(   - \frac{ \Omega_{1}  }{ ( \xi \beta  +  \Omega_{1} ) \xi \sqrt{ \mu_{p} } }    \right)^{  k - 1 }    \biggr]  \cdot  \MeijerG*{4}{5  }{ 5 }{ 5  }{ \mathcal{K}_{3}   }{ \mathcal{K}_{2}  }{    \frac{ \mu_{2} }{ \mu_{1} }  }.
\label{eq:pnzsc3}
\end{gather}
\hrulefill
\vspace*{-12.6pt}
\end{figure*}

\subsection{Asymptotic SOP Analysis \label{subsec:asy_pnzsc}}

\begin{figure*}[t!]
\centering
\begin{minipage}{.5\textwidth}
\centering
  \includegraphics[width=0.96\linewidth,keepaspectratio,angle=0]{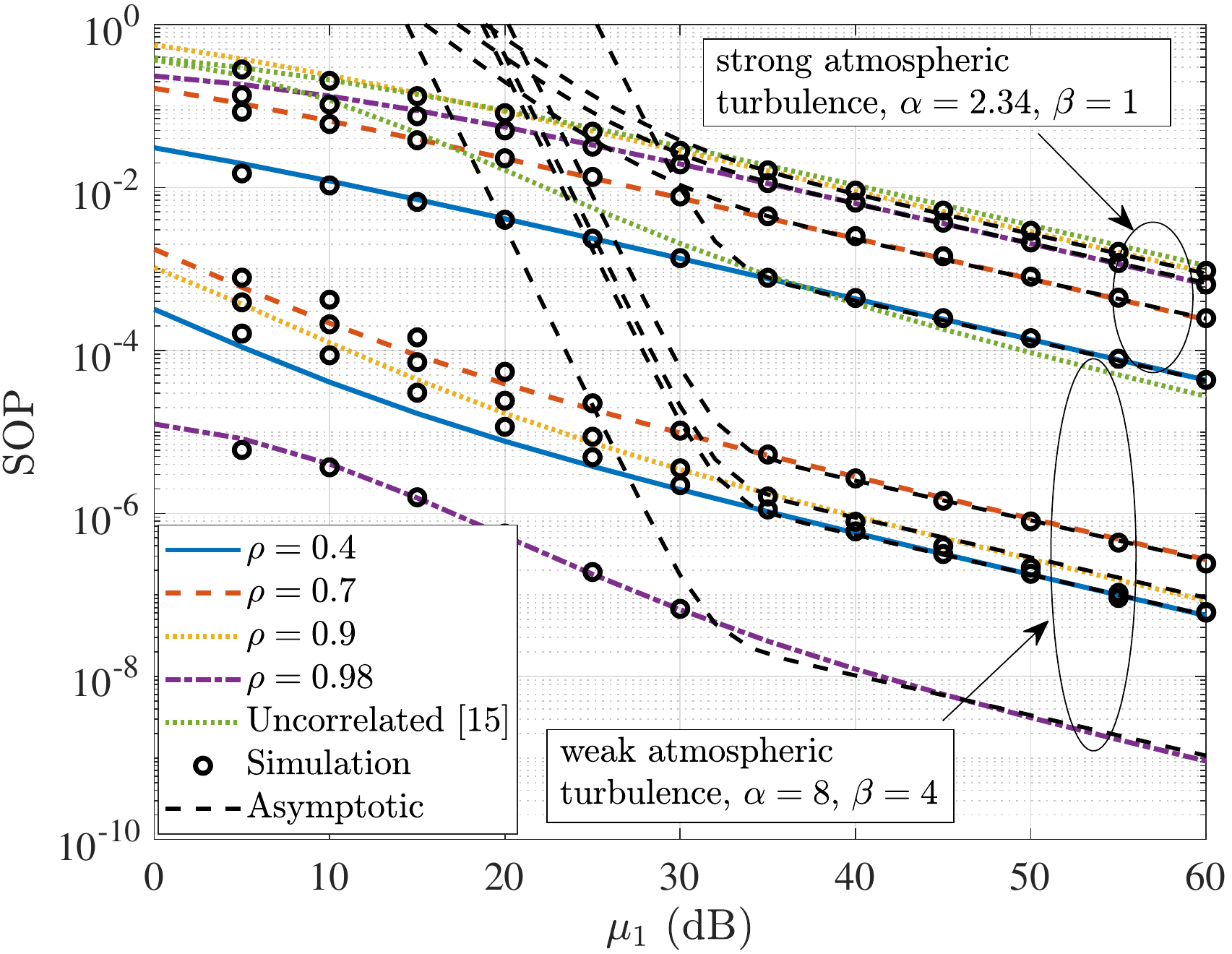}
  \caption{SOP under varying correlations and atmospheric turbulences.}
  \label{fig:sop1}
\end{minipage}\hfill
\begin{minipage}{.5\textwidth}
\centering
  \includegraphics[width=0.96\linewidth,keepaspectratio,angle=0]{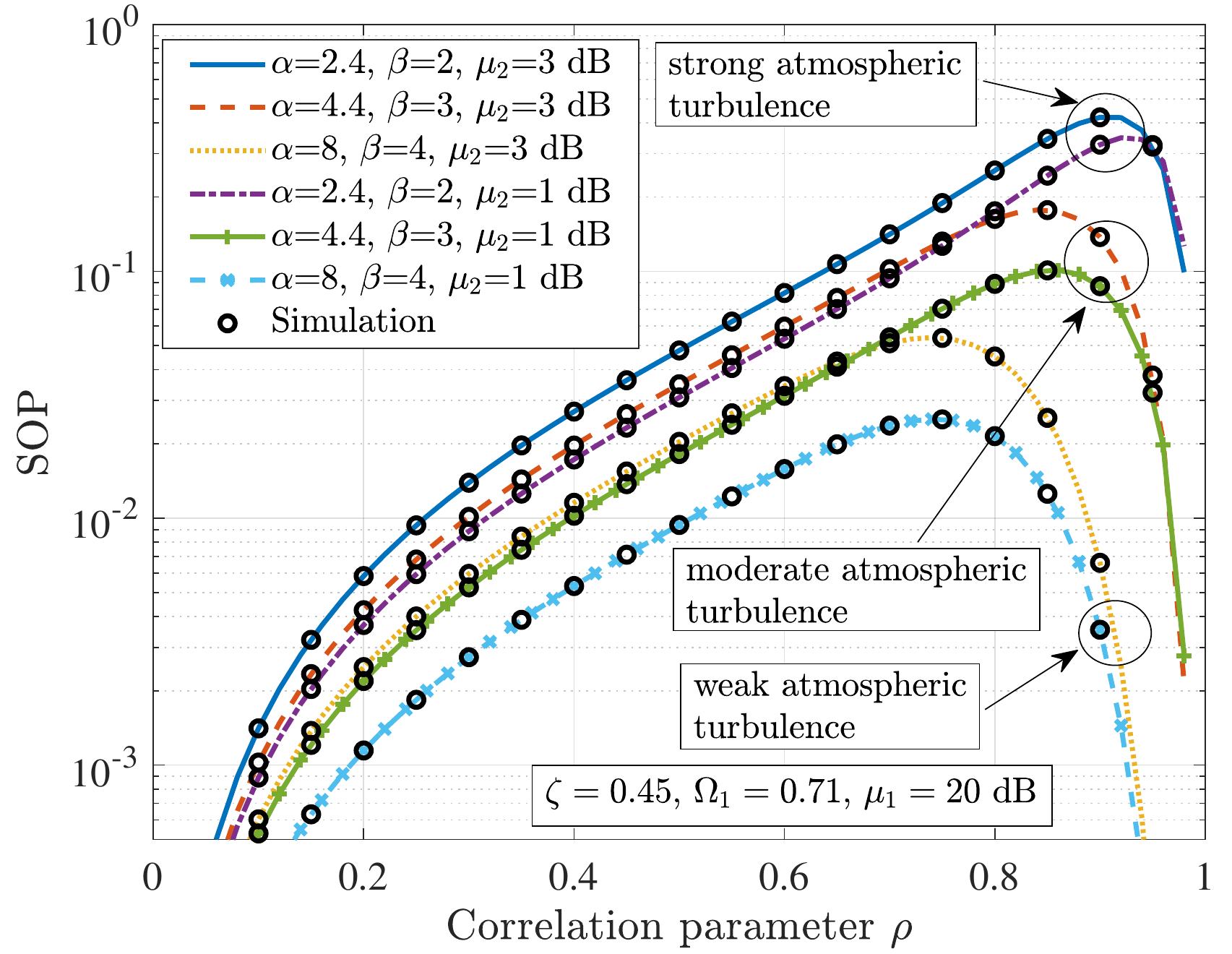}
  \caption{SOP in terms of varying correlation parameter.}
  \label{fig:sop2}
\end{minipage}\hfill
\end{figure*}

\begin{figure*}[t!]
\centering
\begin{minipage}{.5\textwidth}
\centering
  \includegraphics[width=0.96\linewidth,keepaspectratio,angle=0]{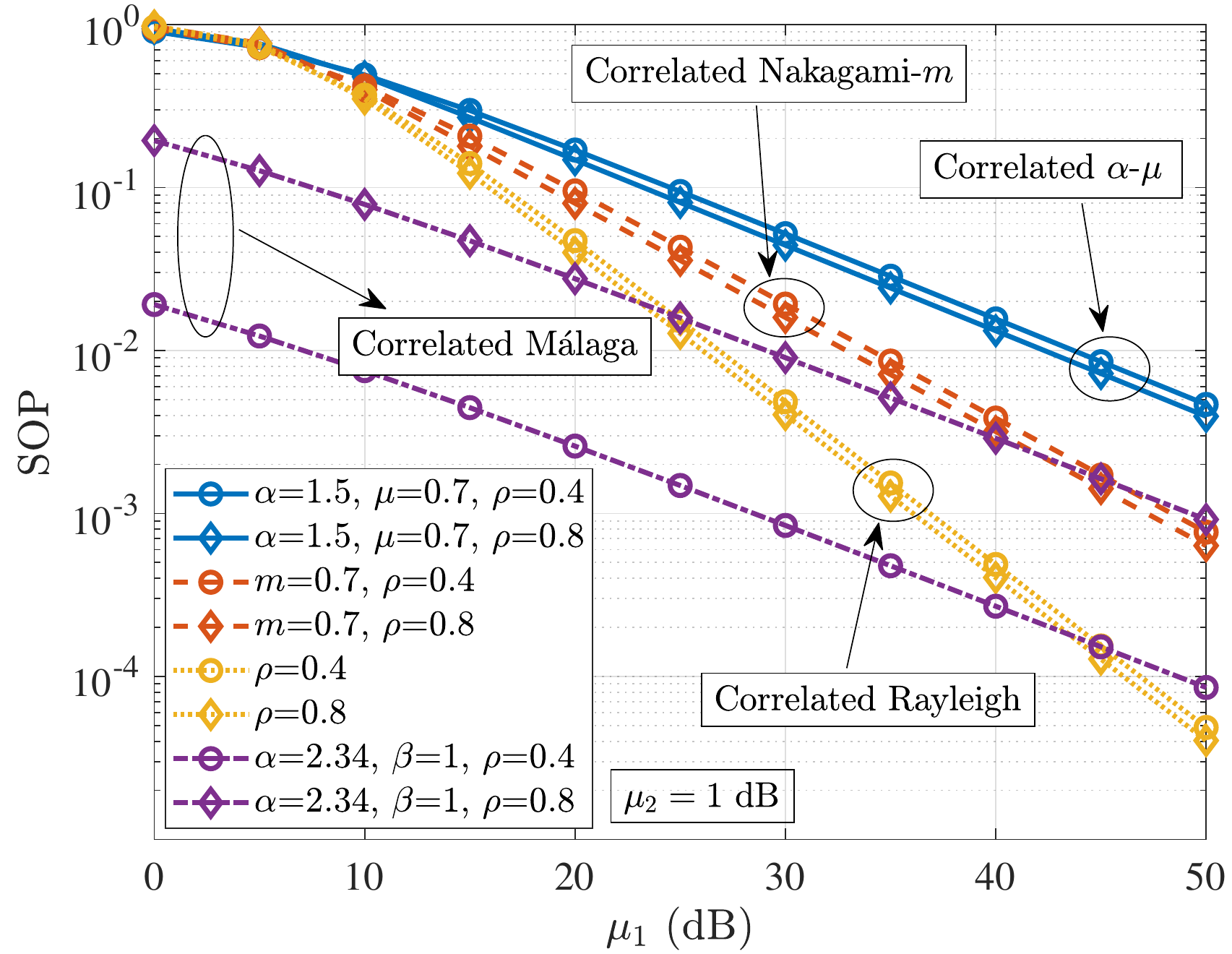}
  \caption{SOP under different correlated RF and FSO fading channels.}
  \label{fig:sop_aux}
\end{minipage}\hfill
\begin{minipage}{.5\textwidth}
\centering
  \includegraphics[width=0.96\linewidth,keepaspectratio,angle=0]{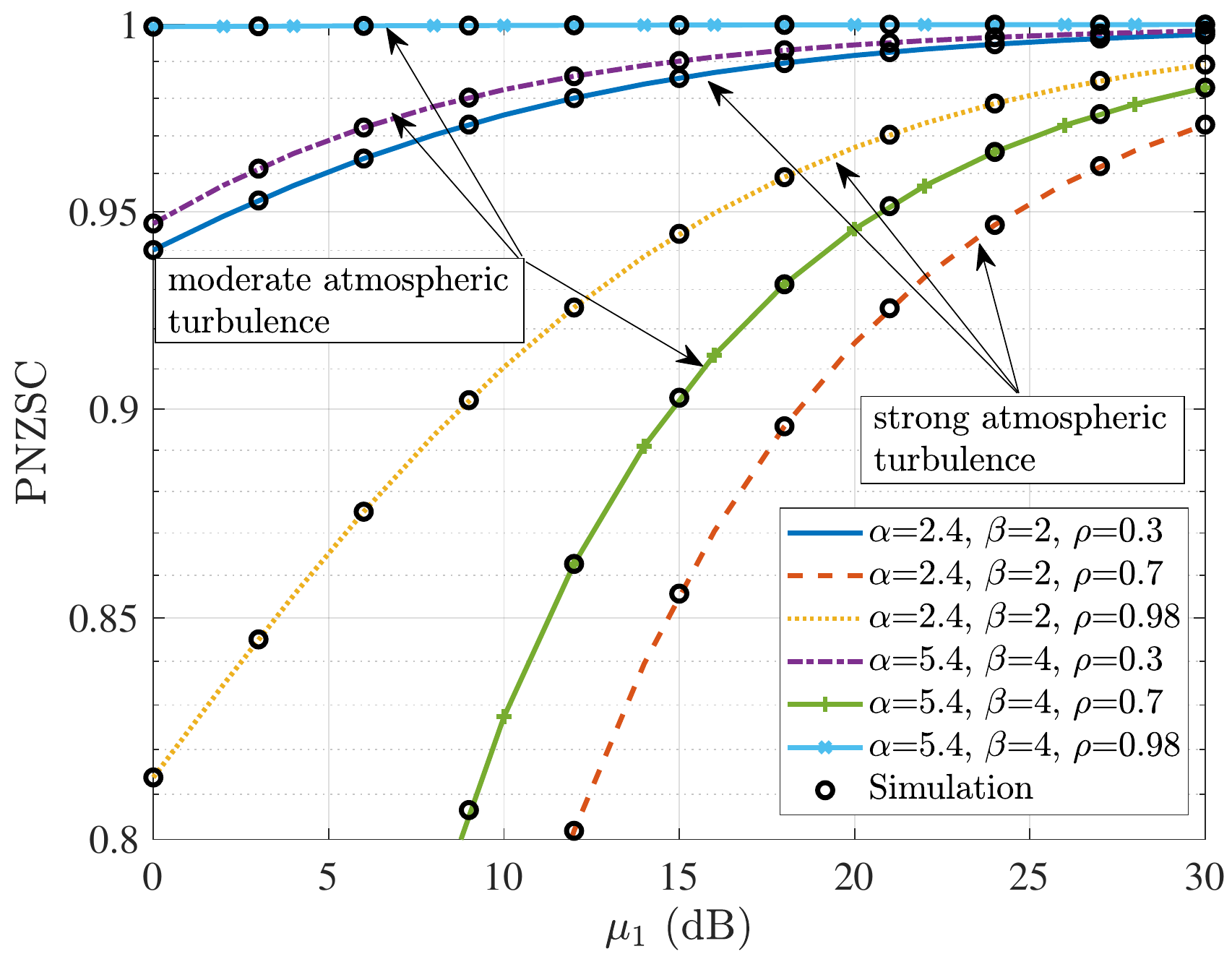}
  \caption{PNZSC under varying correlation and channel conditions.}
  \label{fig:pnzsc}
\end{minipage}\hfill
\end{figure*}

To gain more insights on the impact of link correlation as well as M{\'a}laga fading parameters on the secrecy performance of FSO communication, we conduct secrecy diversity analysis for SOP by considering high values of the average SNR $\mu_1$.

For high values of SNR $\mu_1$, the argument of the Meijer G-function in (\ref{eq:sop4}) tends to 0. Rewriting the function in terms of the generalized hypergeometric function using Slater's theorem \cite{ai2019physicalpj} and applying the relation $\lim\limits_{z\rightarrow 0}{}_{p}F_{q}{(a_p; b_q; \pm z)\rightarrow 1}$, the following asymptotic relation is deduced:
\begin{equation}
\lim_{\mu_1\rightarrow \infty} G_{1,5}^{4,1}\left(\frac{{\mathcal{C}^2}z}{16\mu_1}\left\vert \begin{matrix}
1-\frac{\alpha+t+k}{4}\\
{\mathcal{K}_2}
\end{matrix} \right.\right) \approx \sum_{h=1}^4{\mathcal{D}} \! \cdot \! \Big(\frac{{\mathcal{C}^2}z}{16\mu_1}\Big)^{b_h},
\label{eq:asy_sop1}
\end{equation}
where ${\mathcal{D}}=\frac{\Pi_{g=1}^4\Gamma(b_g-b_h)^*\Gamma(1+b_h-a_1)}{\Gamma{(1+b_h-b_5)}}$, $z=(1+\frac{\mu_2s_{\iota}^8}{16{\mathcal{C}}^2})\Theta-1$, $a_1=1-\frac{\alpha+t+k}{4}$, $b_1=\frac{\alpha+t-k}{4}$, $b_2=\frac{\alpha+t-k+2}{4}$, $b_3=-\frac{\alpha+t-k}{4}$, $b_4=-\frac{\alpha+t-k-2}{4}$, $b_5=-\frac{\alpha+t+k}{4}$; and $(\cdot)^{*}$ indicates to ignore the terms with the subscript $g=h$.

It is obvious that the dominant term in the asymptotic expression for SOP corresponds to the lowest power of the SNR $\mu_1$. Utilizing the asymptotic relation in (\ref{eq:asy_sop1}) for (\ref{eq:sop4}), we have that the lowest power of the SNR $\mu_1$ occurs when $t=0$ and $k=1$. Then, the asymptotic expression for the SOP for large values of the SNR $\mu_{1}$ can be approximated as
\begin{equation}
 P_o\approx \sum_{h=1}^4{\mathcal{E}} \cdot  {\mu_1}^{-\big(b_h+\frac{\alpha+1}{4}\big)},
 \label{eq:asy_sop2}
\end{equation}
where ${\mathcal{E}}=\frac{{\mathcal{D}}\zeta^{\beta-1}\beta^{\alpha+\beta-1}}{2\sqrt{\pi}(\zeta\beta+\Omega_1)^{\alpha+\beta-1}}\frac{(1-\rho^2)^{-1}\mathcal{C}^{-\frac{\alpha+1}{2}}}{[\Gamma(\alpha)]^2\Omega^{\alpha+1}}\sum\limits_{{\iota}=1}^L w_{\iota}s_{\iota}^{4\alpha-1} \cdot z^{b_h+\frac{(\alpha+1)}{4}}  \cdot \big(\frac{{\mathcal{C}}^2}{16}\big)^{b_h} \cdot U\big(\frac{1}{2}+\alpha-1, 2\alpha-1; 2s_{\iota}^2\big)$.

Thus, on substituting the values of $b_h$, $h\in [1,4]$ in (\ref{eq:asy_sop2}), it can be concluded that the asymptotic slope of the SOP curve is $\min\big\{\frac{\alpha}{2}, \frac{1}{2}\big\}$. It is clear that the slope depends only on the fading severity due to atmospheric turbulence and is independent of the correlation condition.

\section{Numerical Results and Discussions \label{sec:results}}

In this section, we numerically evaluate the impact of channel correlation between the main and wiretap links on the secrecy outage performance of FSO communications over M{\'a}laga turbulence channels.

%

In Fig. \ref{fig:sop1}, we compare the SOP of the considered FSO system for different values of correlation parameter $\rho$ as a function of legitimate SNR $\mu_1$ for a fixed eavesdropper SNR $\mu_2 = 5$ dB. The other parameters are $b_0 = 0.423$, $\delta = 0.84$, and $\Omega_1 = 2.04$ [15]. It is observed that the SOP performance improves as the SNR $\mu_1$ increases. For the weak atmospheric turbulence scenario, it is seen that as $\rho$ increases from 0.4 to 0.7, the SOP performance becomes poor. But then as the correlation parameter $\rho$ increases from 0.7 to 0.98, the SOP becomes better. It is widely believed that the correlation degrades the average secrecy capacity performance of wireless communications \cite{mathur2019secrecycorrelated}. However, this is only partially true for the secrecy outage performance of the FSO communications within some range of the correlation parameter according to Fig. \ref{fig:sop1}. In other words, there is an SNR penalty for achieving a target SOP as the correlation increases within some range. For instance, for achieving an SOP of $10^{-3}$, $\mu_1 = 55$ dB is required for strong turbulence with $\rho = 0.7$, while the same is achieved at  $\mu_1 = 45$ dB for $\rho = 0.5$, thereby indicating a power penalty of approximately 10 dB on increasing $\rho$. By comparing with the SOP under uncorrelated channel condition, it is obvious that correlation impacts the SOP significantly compared to the uncorrelated case, showing the importance of taking it into consideration.

Additionally, the asymptotic analysis can also be verified from Fig. \ref{fig:sop1}, i.e, when both SOP and $\mu_{1}$ are written in decibel unit, the linear trend will dominate when the SNR is larger than around 35 dB. To further verify the slope, we consider the curve of weak turbulence with $\rho=0.9$, the SOP is $8.653\times10^{-7}$ for SNR $\mu_{1}$ of 50 dB while it is $2.712\times10^{-7}$ for $\mu_{1}$ of 60 dB. Therefore, the value of the slope is calculated as $ \log_{10}(8.653 \cdot 10^{-7}) - \log_{10}(2.712 \cdot 10^{-7}) = 0.5039 \approx 0.5 = \min\big\{\frac{\alpha}{2}, \frac{1}{2}\big\}$, which verifies the theoretical asymptotic analysis.

The non-monotonic impact of correlation $\rho$ on the SOP as observed in Fig. \ref{fig:sop1} is further delved in Fig. \ref{fig:sop2}, where the SOP is plotted as a function of the correlation coefficient $\rho$ under different atmospheric turbulences. It is observed that as the correlation parameter $\rho$ increases, the SOP performance first degrades, but the SOP performance starts to improve beyond a certain value of $\rho$. This reveals that the correlation between the main and eavesdropper channels can help to improve the secrecy performance of FSO systems. Moreover, the value of the critical $\rho$, i.e., the value beyond which the SOP performance starts improving, reduces as we move from strong turbulence regime to weak turbulence regime.

The results in Figs. \ref{fig:sop1} and \ref{fig:sop2} clearly indicate that the secrecy outage performance exhibits a non-monotonic behavior with the channel correlation between the main and eavesdropping links. The reason for this monotonic behavior is as follows: when the correlation is small, the eavesdropper will take advantage of the similarities due to correlation between the main and eavesdropping channels. However, when the correlation is significantly large enough, the legitimate nodes can gain knowledge on eavesdropper's CSI even under passive eavesdropping thanks to hight correlation. In this sense, there exists a critical value of the correlation parameter $\rho$, where the trend on the impact of correlation on SOP performance reverses. Analytically speaking, the final expression of SOP contains terms of $\rho^2$ and $1-\rho^2$. As $\rho$ increases, terms corresponding to $\rho^2$ increase and  terms pertaining to $1- \rho^2$ decrease. To assert which term dominates, varying values of $\rho$ can be put and it is seen that after the critical value of $\rho$, the terms pertaining to  $1- \rho^2$  begin to dominate and hence the SOP improves again. The non-monotonic impact of channel correlation were also reported in \cite{ferdinand2013physical} and \cite{liu2013outage}. However, it is obvious that the correlation for the M{\'a}laga channels impact the secrecy outage performance more significantly than Nakagami-$m$ and lognormal distributions studied in \cite{ferdinand2013physical} and  \cite{liu2013outage}. This observation is verified in Fig. \ref{fig:sop_aux}, which shows SOP under different correlated RF and FSO fading channels.

In Fig. \ref{fig:pnzsc}, we plot PNZSC as a function of the SNR $\mu_1$ for varying values of correlation coefficient  $\rho$ under strong and moderate turbulence scenarios. It is seen that the PNZSC improves as $\mu_{1}$ increases for a given atmospheric turbulence and correlation coefficient. As $\rho$ increases, the PNZSC performance degrades but on further increasing $\rho$ beyond a critical value, the PNZSC performance improves again. This non-monotonic impact of correlation on PNZSC is in accordance with the trend for SOP.

\section{Conclusion \label{sec:conclusion}}

In this paper, we studied the secrecy outage performance of FSO communications, where the main and wiretap links experience arbitrarily correlated M{\'a}laga fading. Novel expressions for SOP were derived, and asymptotic analysis on the SOP was also conducted. The obtained results provide useful insights on practical scenario of FSO communication security. The main findings of the paper are as follows:  (\romannum{1}) Counterintuitively, the secrecy outage performance demonstrates a non-monotonic behavior with the increase of correlation. In other words, the correlation can be exploited to enhance the SOP performance; (\romannum{2}) The critical value of correlation parameter (i.e., the value beyond which increased correlation indicated better SOP performance) decreases as the atmospheric turbulence varies from strong to weak conditions; and (\romannum{3}) The asymptotic slope of the SOP is independent of correlation and depends only on the fading severity due to atmospheric turbulence.

%
%
%
%

%

\ifCLASSOPTIONcaptionsoff
  \newpage
\fi


\balance
\bibliographystyle{IEEEtran}
\bibliography{PLS_correlated_malaga_fading_wcl_ref}

%
%


\end{document}